\documentclass[conference]{IEEEtran}
\usepackage{amssymb}
\usepackage{mathtools}
\usepackage{graphicx}
\graphicspath{ {Figures/} }
\usepackage{xr}
\usepackage{booktabs}
\usepackage{cleveref}

\newcommand{\Lagr}{\mathcal{L}}
\DeclarePairedDelimiter\abs{\lvert}{\rvert}
\makeatletter
\let\oldabs\abs
\def\abs{\@ifstar{\oldabs}{\oldabs*}}

\usepackage{cite}

\ifCLASSINFOpdf
\else
\fi
\begin{document}

\title{Classifying HCP Task-fMRI Networks Using\\Heat Kernels}

%
\author{\IEEEauthorblockN{Ai Wern Chung\IEEEauthorrefmark{1},
Emanuele Pesce\IEEEauthorrefmark{2},
Ricardo Pio Monti\IEEEauthorrefmark{1} and
Giovanni Montana\IEEEauthorrefmark{1}}
\IEEEauthorblockA{\IEEEauthorrefmark{1}Department of Biomedical Engineering, Division of Imaging Sciences \& Biomedical Engineering,\\King's College London, UK}
\IEEEauthorblockA{\IEEEauthorrefmark{2}Department of Mathematics, Imperial College London, UK}}


\maketitle

\begin{abstract}

Network theory provides a principled abstraction of the human brain: reducing a complex system into a simpler representation from which to investigate brain organisation. Recent advancement in the neuroimaging field are towards representing brain connectivity as a dynamic process in order to gain a deeper understanding of the interplay between functional modules for efficient information transport. In this work, we employ heat kernels to model the process of energy diffusion in functional networks. We extract node-based, multi-scale features which describe the propagation of heat over 'time' which not only inform the importance of a node in the graph, but also incorporate local and global information of the underlying geometry of the network. As a proof-of-concept, we test the efficacy of two heat kernel features for discriminating between motor and working memory functional networks from the Human Connectome Project. For comparison, we also classified task networks using traditional network metrics which similarly provide rankings of node importance. In addition, a variant of the Smooth Incremental Graphical Lasso Estimation algorithm was used to estimate non-sparse, precision matrices to account for non-stationarity in the time series. We illustrate differences in heat kernel features between tasks, and also between regions of the brain. Using a random forest classifier, we showed heat kernel metrics to capture intrinsic properties of functional networks that serve well as features for task classification.

\end{abstract}

\IEEEpeerreviewmaketitle

\section{Introduction}

Network theory provides a simple abstraction of neural connectivity in the human brain and its use to investigate brain organisation in neuroimaging has gained momentum in recent years. The brain is a complex, interconnected system which maintains both functional segregation and integration designed to optimise information transport~\cite{bullmore_complex_2009}. Due to the integrative nature of brain functionality, there is increasing evidence of cognition being supported by coordinated activity between different functional modules~\cite{fornito_competitive_2012, hellyer_control_2014}. As such, a number of recent studies have sought to capture such dynamic processes using, for example, models of perturbation or spread through a network~\cite{misic_cooperative_2015, stam_relation_2015, gu_controllability_2015}. In this work, we propose the use of heat kernels, a diffusion model, to capture information propagation dynamically through functional brain networks.
 
The heat kernel describes the effect of applying a heat source to a network and observing the diffusion process over 'time'. It encodes the distribution of energy over a network and characterises the underlying structure of the graph. Heat kernels have found success in the field of 3D object recognition~\cite{sun_concise_2009,fang_temperature_2011} where node-based features such as the \textit{heat kernel signature}, HKS~\cite{sun_concise_2009, bronstein_scale-invariant_2010}, and the \textit{average temperature function}, AVG~\cite{fang_temperature_2011}, fair well as shape descriptors. These metrics are able to organise the intrinsic geometry of the network over multiple-scales, capturing local and global 'shapes' in relation to a node via a time parameter. In addition, they are stable and invariant to isometric deformations, and thus are useful for matching objects in different poses. This stability against noise is an attractive property to capture potentially informative features in functional neuroimaging data. As such, these features also incorporate a concept of the most influential nodes as measured by heat propagation in a network. Given the shifting states of neural systems to meet the demands of cognitive function~\cite{mattar_functional_2015}, the use of dynamic, node-based features to investigate brain regions integral for energy transport may be beneficial towards understanding how these states interact with one another~\cite{gu_controllability_2015}. For this reason, we sought to investigate the effectiveness of HKS and AVG as dynamic features for discriminating between task fMRI networks.

For comparison, we also classify the networks using other traditional network measures which provide a ranking of node importance. Centrality measures fulfil such a need of which there are many variants. Two common kinds used in neuroimaging are betweenness~\cite{freeman_set_1977} and eigenvector centrality~\cite{bonacich_power_1987}.
Other measures such as PageRank~\cite{page_pagerank_1999} and HITS~\cite{kleinberg_authoritative_1999} which analyse incoming links in a network can be used to order nodes by influence and are commonly used to rank web pages. The spectral nature of eigenvector, PageRank and HITS mean they capture the influence of a node that extend to global features of the network. As such, we opt to compare these with the heat kernel measures. We also include betweenness centrality as it is commonly used in neuroimaging network analysis.

A distinct challenge in the study of brain networks is addressing potential non-stationarity. While traditionally brain networks had been assumed to be stationary, recent results suggest connectivity between brain regions is high non-stationary. This is particularly true in the context of task based paradigms, such as the Human Connectome Project (HCP) data used in this work. As a consequence, several 
novel methodologies have been proposed to accurately estimate the dynamic properties of functional networks. A common denominator in many of these approaches is the use of regularisation to reduce the number of free parameters. The choice of regularisation will typically depend on the objectives of the proposed algorithm. For example, one of the objectives of the recently proposed SINGLE algorithm \cite{monti_estimating_2014} was to estimate sparse connectivity networks. This naturally led to the inclusion of $\ell_1$ regularisation. With the heat kernel, we are interested in capturing energy flow over an entire network whereby having a denser network would enable this and may also contain additional information on functional modules. Furthermore, the stable property of heat kernel features may be able to account for any noise in the estimated networks that is not a feature the task. Here, we use a variant of SINGLE to estimate non-sparse, precision matrices that were centred on the task's peak response to account for non-stationarity in the time series.

We present a preliminary investigation into the efficacy of node features derived from the heat kernel to classify between two task-evoked connectivity networks. As comparison, we also classify using a set of centrality measures which capture a notion of network topology similar to these heat kernel features.

The remainder of the paper is organised as follows: in Section II we first describe the non-sparse variant of the SINGLE algorithm, followed by details of heat kernel and centrality features and the classification method. The results are reported in Section III, before concluding in Section IV.

\section{Methods}

\subsection{Human Connectome Project data}
We used a total of 491 subjects from the HCP 500 subject release~\cite{van_essen_human_2012}. We chose motor and working-memory (WM) 2-back tasks, RL phase-encoding acquisitions. Cerebral spinal fluid, white matter and motion were regressed from the time series. A total of $V = 84$ (nodes) regions comprised of 68 cortical and 16 subcortical structures were defined from the Desikan-Killiany and ASEG atlases, respectively. Signals were extracted and averaged within each of these regions. The time of maximal peak, $peak_t$, in the haemodynamic response of a representative subject was identified for each task. A connectivity matrix was then estimated from the time series over a localised temporal window centred on $peak_t$ for all subjects.

\subsection{Estimation of time-varying networks}

We estimated a non-sparse, precision network using a modified version of the SINGLE algorithm~\cite{monti_estimating_2014} in which $\ell_2$ regularisation was employed. Such an approach retains many of the advantages of the original SINGLE algorithm, such as leveraging information across chronologically proximal networks, while also yielding dense networks for our objective. Briefly, the SINGLE algorithm produces time-varying connectivity networks by estimating a corresponding precision matrix at each observation. Thus for the $k$th subject we obtain a sequence of precision matrices $\{ \Theta^{(k)}_i \}_{i=1}^T$. $\Theta^{(k)}_i$ encodes the partial correlation structure at the $i$th time point. It follows that each precision matrix is associated with a weighted graph, $G^{(k)}_i$.

The proposed network estimation algorithm looks to solve the following convex optimisation problem:
\begin{equation}
\label{SINGLE_cost}
 \{\hat \Theta_i^{(s)}\} = \underset{\Theta_i^{(s)}}{\mbox{argmin}} \left  \{f(\{\Theta_i^{(s)}\}) + g_{\lambda_1, \lambda_2}(\{\Theta_i^{(s)}\})  \right \}.
\end{equation}
Here $f(\{\Theta_i^{(s)}\})$ 
is a negative log-likelihood term
and $g_{\lambda_1, \lambda_2}(\{\Theta_i^{(s)} \})$ is a regularisation term consisting of 
$\ell_2$ penalty terms:
$$g_{\lambda_1, \lambda_2}(\{\Theta_i^{(s)} \}) = \lambda_1 \sum_{i=1}^T || \Theta_i^{(s)}||_2^2 + \lambda_2 \sum_{i=2}^T || \Theta_i^{(s)} -  \Theta_{i-1}^{(s)}||_2^2. $$ 

The adjusted algorithm required the selection of three parameters. The first is a Gaussian kernel bandwidth, $h$, which governs the portion of the time series centred on $i$ from which to estimate sample covariance matrices. The use of a Gaussian kernel ensures that observations are weighted according to their proximity to $i$. The remaining parameters are regularisation parameters $\lambda_1$ and $\lambda_2$. Following~\cite{monti_graph_2015}, the choice of kernel bandwidth was selected via cross-validation over a random subset of ten subjects. This resulted in  $h=12.5$ and $h=17.5$ for the motor and WM tasks respectively. For the work presented here, we use $\Theta^{(k)}_i$ from a single observation at $i$ = $peak_{t}$ for each task. Let $\Theta^{(k)}_i$ at $i = peak_t$ be defined as $G$ from now.

\subsection{Heat kernel features}

$G = (V,E)$ where $V$ is the set of $|V|$ nodes on which the graph is defined and $E \subseteq V\times V$ the corresponding set of edges. A weighted matrix, $W$, is $W(u,v) = w_{uv}$ where $w_{uv}$ is the corresponding edge strength. A diagonal strength matrix, D, is $D(u,u) = deg(u) = \sum_{v \in V}w_{uv}$. The Laplacian of $G$ is $\Lagr = D - W$, and the normalised Laplacian is given by $\hat\Lagr = D^{-1/2}\Lagr D^{-1/2}$.

The heat kernel, $H(t)$, is the fundamental solution to the standard, partial differential equation of a diffusion process, 
$$\frac{\partial H(t)}{\partial t} = -\hat\Lagr H(t),$$ 

and can be computed analytically: $$H(t) = \text{exp}[-t\hat\Lagr].$$

$H(t)$ describes the flow of energy through $G$ at time $t$ where the rate of flow is governed by $\hat\Lagr$ calculated from $\abs{W}$. $H(t)$ is a symmetric $|V| \times |V|$ matrix where the entry $H(t)_{u,v}$ represents the amount of heat transfer between nodes $u$ and $v$ after time $t$. The heat kernel average temperature function, $$AVG(t)_{u} = \frac{1}{N-1}\sum_{v, v\ne u}H(t)_{u,v},$$ evaluates the importance of $u$ whereby a strongly connected node would have edges with high energy transfer~\cite{fang_temperature_2011}. The heat kernel signature, $$HKS(t)_{u} = H(t)_{u,u},$$ captures the intrinsic geometry of the network for node $u$ at time $t$~\cite{sun_concise_2009}. Both HKS and AVG report slightly different aspects of heat propagation in a network. As the diagonal of the heat kernal matrix, $HKS(t)_{u,u}$ is indicative of the amount of energy stored in a node at any given time. $AVG(t)_{u,u}$ is indicative of the importance of node $u$ by capturing the average heat distributed between it and every other node in the network. This is because an 'edge' entry in $H(t)_{u,v}$ contains the amount of heat transported between nodes $u$ and $v$ via all possible pathways. For a given $G$, we calculated 300 $H(t)$ over a range of $t=[0.05,0,1, ..., 15.0]$ and normalised each kernel within $t$. Both HKS and AVG were extracted from all nodes in each heat kernel. This gave 25,200 heat kernel features per each subject per task.

\subsection{Network centrality measures}

As a comparison, several node-based measures which provide information on node rank were extracted from $G$. These were betweenness, eigenvector centralities, PageRank and HITS (as implemented in NetworkX~\cite{hagberg_exploring_2008}). Briefly, betweenness centrality ranks node importance by the number of shortest paths which traverse through a node. Eigenvector centrality determines node influence by not only considering its association with other nodes, but also takes into account the importance of its neighbouring nodes. PageRank and HITS are link analysis algorithms, typically used by search engines to rank pages from a search result. PageRank is closely related to eigenvector centrality but differs by incorporating a damping factor on contributing neighbours. HITS captures relational properties between nodes by scoring them as hub (important for pointing to many other nodes) or authority (important as it is often pointed to by many hubs) entities. Given the symmetry of our connectivity matrices, hubs and authority scores from HITS are equivalent. Each centrality measure resulted in a vector of 84 features.

\subsection{Networks classification}

Our choice of algorithm to classify between motor and WM features was a random forest~\cite{breiman_random_2001}. Our choice for the random forest was its ability to rank features by Gini importance for the classification process. We used the scikit-learn implementation~\cite{pedregosa_scikit-learn:_2011}, in with 1000 estimators (trees). Each separate feature was vectorised (e.g. vector length 25,200 for HSK, or a vector of 84 for HITS) as inputs into the classifier. A stratified, 10-fold cross validation approach was adopted to evaluate the performance of the six node-based features. Classification performance scores of accuracy, sensitivity and specificity were recorded and averaged across all folds. 

\section{Results}

\begin{figure}[!t]
\centering
\includegraphics[width=3.5in]{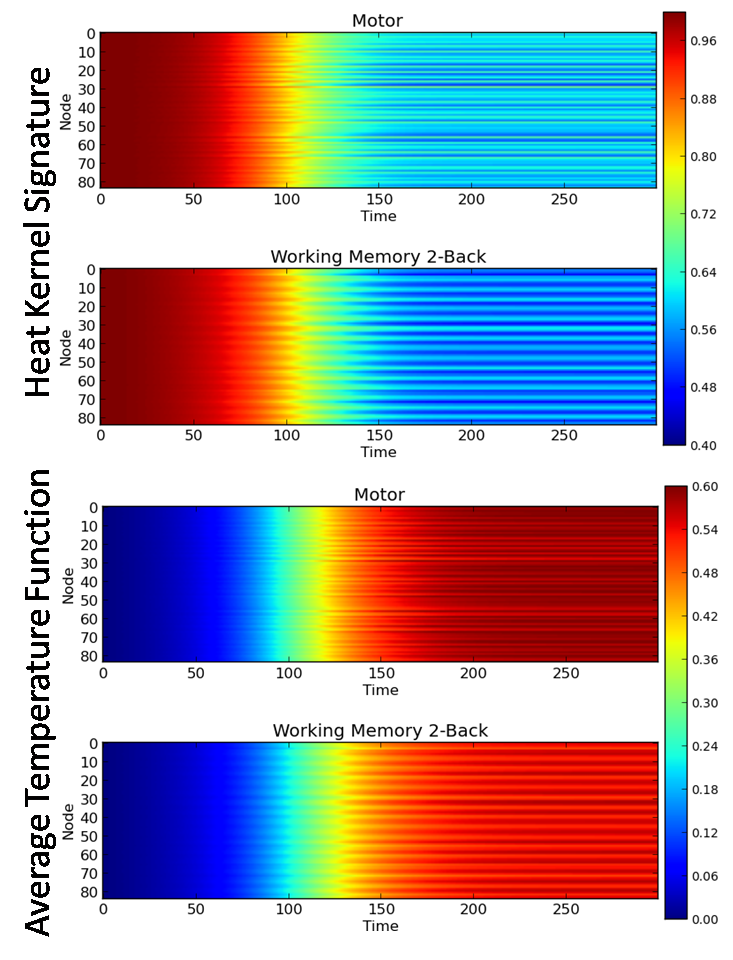}
\caption{Node heat transition maps showing the change in HKS and AVG for each node with increase $t$. Each map is an average across all subjects for motor and WM task networks.}
\label{fig:heatmaps}
\end{figure}

\begin{figure}[!t]
\centering
\includegraphics[width=3in]{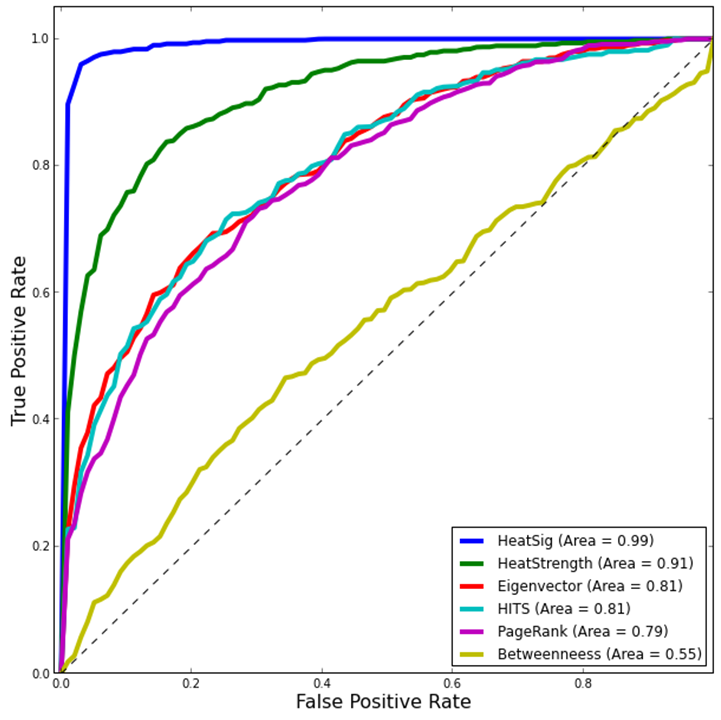}
\caption{ROC curves for the 6 node-feature tested to classify task-fMRI networks. The area under the curves are reported.}
\label{fig:ROCplot}
\end{figure}

\begin{figure}[!t]
\centering
\includegraphics[width=3.5in]{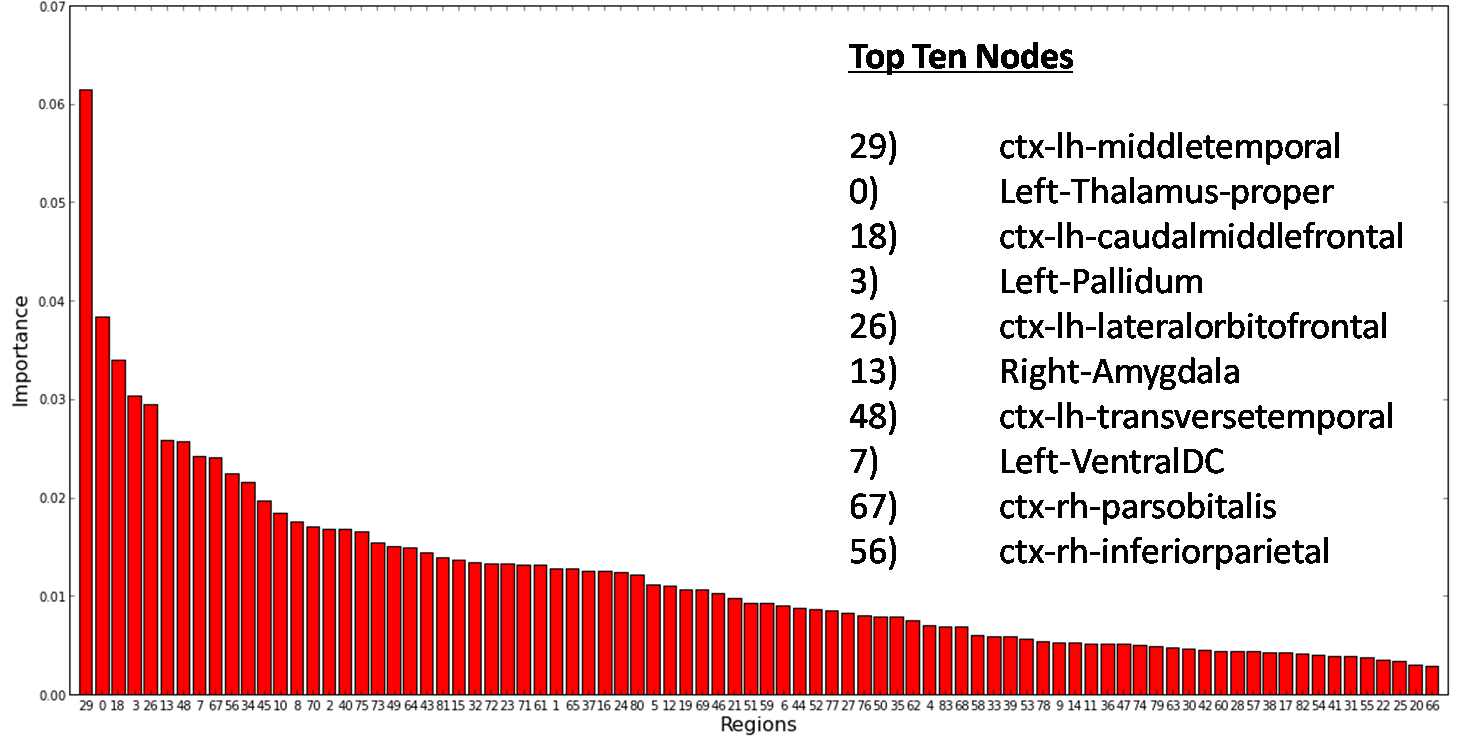}
\caption{Random forest leaves ordered by importance from a single cross-validation run classifying task networks by HKS. Bar plots are the ranking for each network node summed across all HKS over $t$.}
\label{fig:barchart}
\end{figure}

Group averaged heat transition maps qualitatively reveal differences in heat kernel features between motor and WM task networks, Fig.~\ref{fig:heatmaps}. With increasing time, there is an exponential decrease in HKS in all brain regions. For AVG, the opposite occurs; it increases with time. In addition, variations in the rate of change can be seen both between task networks and between regions within the networks for both metrics. ROC plots in Fig.~\ref{fig:ROCplot} show heat kernel features had the largest areas under their curves, outperforming all remaining graph measures tested. Average performance accuracy across all folds were: HKS$=95.9\%$, AVG$=83.7\%$, eigenvector$=73.5\%$, HITS$=73.0\%$, PageRank$=69.4\%$, betweenness$=53.8\%$. Given the classification performance in Fig.~\ref{fig:ROCplot}, the features can be broadly categorised into three groups: heat kernel features (HKS, AVG), spectral-based features (eigenvector, HITS, PageRank) and path-based features (betweenness). The most discriminative nodes for classifying between task networks ranked in decreasing order of importance for a single cross-validation fold is plotted in Fig.~\ref{fig:barchart}.  

\section{Conclusion}

We presented exploratory work on the use of node-based heat kernel features applied to functional MRI data. We illustrated their ability to represent functional connectivity as a dynamic process and their use for discriminating between two different task-evoked networks.

Traditional network measures which rely on paths and path lengths may not be suitable for describing network topology in functional graphs. Such measures are more appropriate for networks based on physical, anatomical connections whereas an interpretation of functional connectivity may be better served by a diffusion model such as the heat kernel. Heat kernel metrics also have the advantage of possessing a multi-scale property where by varying $t$, $HKS(t)_u$, for example, contains information of the network's geometry as defined within the local or global area surrounding $u$. As $t$ increases, HKS at $u$ decreases as it represents the average behaviour over an ever increasing region surrounding $u$, and thus eventually stabilises. We have shown this to be the case in functional networks, and also revealed variations in the rate of change between brain regions. AVG on the other hand increases with $t$. It may be that over time, it measures the heat dispersed from nodes into the system and the more influential the node, the more energy that was distributed is associated with it. We demonstrated that these changes in heat propagation are useful features for discriminating between functional task networks with high levels of accuracy when compared with several other equivalent centrality measures.

It is quite possible that the most discriminative nodes are not associated with the tasks tested. Such brain regions may differ between tasks, but in themselves, are not significant features of the task. In addition, performance of the WM task involves a motor response, thus there will be similar activations in both networks.

Future work will attempt to investigate the serial correlation between heat kernel features per node evident in the heat maps. In addition, the advantage of SINGLE to address non-stationarity in fMRI time series can be exploited by using all networks computed temporally. Heat kernel features could then be extracted from all these networks and analysed through the evolution of a task.

\bibliography{PRNI_ref2}

\end{document}